\documentclass[12pt]{article}
\usepackage{amsmath,amsfonts,amssymb}

\begin{document}
\thispagestyle{empty}

\def\theequation{\arabic{section}.\arabic{equation}}
\def\a{\alpha}
\def\b{\beta}
\def\g{\gamma}
\def\d{\delta}
\def\dd{\rm d}
\def\e{\epsilon}
\def\ve{\varepsilon}
\def\z{\zeta}
\def\B{\mbox{\bf B}}

\newcommand{\h}{\hspace{0.5cm}}

\begin{titlepage}
\renewcommand{\thefootnote}{\fnsymbol{footnote}}
\begin{center}
{\Large \bf M2-brane Perspective on }
\vskip .3cm
{\Large \bf ${\cal N}=6$ Super Chern-Simons
Theory at Level $k$}
\end{center}
\vskip 1.2cm \centerline{\bf Changrim  Ahn and P. Bozhilov
\footnote{On leave from Institute for Nuclear Research and Nuclear
Energy, Bulgarian Academy of Sciences, Bulgaria.}}

\vskip 10mm

\centerline{\sl Department of Physics} \centerline{\sl Ewha Womans
University} \centerline{\sl DaeHyun 11-1, Seoul 120-750, S. Korea}
\vspace*{0.6cm} \centerline{\tt ahn@ewha.ac.kr,
bozhilov@inrne.bas.bg}

\vskip 20mm

\baselineskip 18pt

\begin{center}
{\bf Abstract}
\end{center}
\h Recently, O. Aharony, O. Bergman, D. L. Jafferis and J. Maldacena
(ABJM) proposed three-dimensional super Chern-Simons-matter theory,
which at level $k$ is supposed to describe the low energy limit of
$N$ M2-branes. For large $N$ and $k$, but fixed 't Hooft coupling
$\lambda=N/k$, it is dual to type IIA string theory on $AdS_4\times
\mathbb{CP}^3$. For large $N$ but {\it finite} $k$, it is dual to M
theory on $AdS_4\times S^7/Z_k$. In this paper, relying on the
second duality, we find exact giant magnon and single spike
solutions of membrane configurations on $AdS_4\times S^7/Z_k$
by reducing the system to the Neumann-Rosochatius integrable model.
We derive the dispersion relations and their finite-size corrections
with explicit dependence on the level $k$.

\end{titlepage}
\newpage
\baselineskip 18pt

\def\nn{\nonumber}
\def\tr{{\rm tr}\,}
\def\p{\partial}
\newcommand{\bea}{\begin{eqnarray}}
\newcommand{\eea}{\end{eqnarray}}
\newcommand{\bde}{{\bf e}}
\renewcommand{\thefootnote}{\fnsymbol{footnote}}
\newcommand{\be}{\begin{equation}}
\newcommand{\ee}{\end{equation}}

\vskip 0cm

\renewcommand{\thefootnote}{\arabic{footnote}}
\setcounter{footnote}{0}

\setcounter{equation}{0}
\section{Introduction}
The $AdS/CFT$ duality conjecture \cite{AdS/CFT,GKP98,EW98}, has led
to many interesting developments in understanding the correspondence
between type IIB string theory on $AdS_5\times S^5$ and ${\cal N}=4$
super Yang-Mills theory in four dimensions. Recently, new exciting
field for investigations appeared, after the discovery of the new
$AdS_4/CFT_3$ duality \cite{CFT3}. The most promising candidate for description
of this correspondence is the ${\cal N}=6$ super Chern-Simons-matter
theory proposed by ABJM in \cite{ABJM0806}. This theory at level $k$
describes the low energy limit of $N$ M2-branes probing a $C^4/Z_k$
singularity. At large $N$, it is dual to M theory on  $AdS_4\times
S^7/Z_k$. For large $N$ and fixed ratio $N/k$, it also has a 't
Hooft limit, which is dual to type IIA string theory on $AdS_4\times
{\mathbb {CP}^3}$.

After the appearance of \cite{ABJM0806}, many related papers quickly
followed \cite{06.1519}-\cite{10.1246}, investigating different
aspects of the ABJM theory. We would like to mention the discovery
of the exact $S$-matrix, including the dressing phase
\cite{07.1924}, confirming the all-loop Bethe ansatz equations,
conjectured in \cite{07.0777}.

In this article, we propose an M2-brane viewpoint on the ABJM theory
at finite level $k$, by considering membrane configurations with two
angular momenta on $R_t\times S^7/Z_k$ subspace of $AdS_4\times
S^7/Z_k$ background, which exhibit similar properties of giant magnon (GM) \cite{HM06},
dyonic GM \cite{ND} and single spike (SS) \cite{IK07} in string theory.

The paper is organized as follows. In section 2, we find an
appropriate M2-brane embedding into  $R_t\times S^7/Z_k$. In section
3, we show that there exists unique Neumann-Rosochatius (NR)
integrable system, which describes membrane configurations with two
angular momenta, for any finite value of the level $k$. In section
4, based on this NR approach, we give the corresponding M2-brane
GM and SS solutions and the semiclassical energy-charge relations, including
the finite-size effects. We conclude the paper with some comments in section 5.

\setcounter{equation}{0}
\section{Membranes on $AdS_4\times S^7/Z_k$}
Let us start with the following membrane action \bea\label{oma}
S=\int
d^{3}\xi\left\{\frac{1}{4\lambda^0}\Bigl[G_{00}-2\lambda^{j}G_{0j}+\lambda^{i}
\lambda^{j}G_{ij}-\left(2\lambda^0T_2\right)^2\det G_{ij}\Bigr] +
T_2 C_{012}\right\},\eea where \bea\nn &&G_{mn}= g_{MN}(X)\p_m
X^M\p_n
X^N,\h C_{012}= c_{MNP}(X)\p_{0}X^{M}\p_{1}X^{N}\p_{2}X^{P}, \\
\nn &&\p_m=\p/\p\xi^m,\h m = (0,i) = (0,1,2),\\ \nn
&&(\xi^0,\xi^1,\xi^2)=(\tau,\sigma_1,\sigma_2),\h M =
(0,1,\ldots,10),\eea are the fields induced on the membrane
worldvolume from the background metric $g_{MN}$ and the background
3-form gauge field $c_{MNP}$, $\lambda^m$ are Lagrange multipliers,
$x^M=X^M(\xi)$ are the membrane embedding coordinates, and $T_2$ is
its tension. As shown in \cite{NPB656}, the above action is
classically equivalent to the Nambu-Goto and Polyakov  type actions.
In addition, the action (\ref{oma}) gives a {\it unified}
description for the tensile and tensionless membranes.

The equations of motion for the Lagrange multipliers $\lambda^{m}$
generate the independent constraints only \bea\label{M00}
&&G_{00}-2\lambda^{j}G_{0j}+\lambda^{i}\lambda^{j}G_{ij}
+\left(2\lambda^0T_2\right)^2\det G_{ij}=0,\\
\label{0j} &&G_{0j}-\lambda^{i}G_{ij}=0.\eea
Further on, we will work in diagonal worldvolume gauge
$\lambda^{i}=0$, in which the action (\ref{oma}) and the
constraints (\ref{M00}), (\ref{0j}) simplify to \bea\label{omagf}
&&S_{M}=\int d^{3}\xi \mathcal{L}_{M}= \int
d^{3}\xi\left\{\frac{1}{4\lambda^0}\Bigl[G_{00}-\left(2\lambda^0T_2\right)^2\det
G_{ij}\Bigr] + T_2 C_{012}\right\},
\\ \label{00gf} &&G_{00}+\left(2\lambda^0T_2\right)^2\det G_{ij}=0,
\\ \label{0igf} &&G_{0i}=0.\eea

Let us introduce the following complex coordinates on the $S^7/Z_k$
subspace \bea\nn &&z_1=\cos\psi\cos\frac{\theta_1}{2}
e^{i\left[\frac{\varphi}{k}+\frac{1}{2}\left(\phi_1+\phi_3\right)\right]},
\h z_2=\cos\psi\sin\frac{\theta_1}{2}
e^{i\left[\frac{\varphi}{k}-\frac{1}{2}\left(\phi_1-\phi_3\right)\right]},
\\ \nn &&z_3=\sin\psi\cos\frac{\theta_2}{2}
e^{i\left[\frac{\varphi}{k}+\frac{1}{2}\left(\phi_2-\phi_3\right)\right]},\h
z_4=\sin\psi\sin\frac{\theta_2}{2}
e^{i\left[\frac{\varphi}{k}-\frac{1}{2}\left(\phi_2+\phi_3\right)\right]}.
\eea Obviously, they satisfy the relation \bea\nn
\sum_{a=1}^{4}z_a\bar{z}_a\equiv 1.\eea Next, we compute the metric
\bea\nn ds^2_{S^7/Z_k}=\sum_{a=1}^{4}dz_a
d\bar{z}_a=\frac{1}{k^2}\left(d\varphi+kA_1\right)^2+ds^2_{\mathbb
{CP}^3},\eea where \bea\nn
&&A_1=\frac{1}{2}\left[\cos^2\psi\cos\theta_1
d\phi_1+\sin^2\psi\cos\theta_2 d\phi_2 +
\left(\cos^2\psi-\sin^2\psi\right)d\phi_3\right], \\ \nn
&&ds^2_{\mathbb {CP}^3}=
d\psi^2+\sin^2\psi\cos^2\psi\left(\frac{1}{2}\cos\theta_1
d\phi_1-\frac{1}{2}\cos\theta_2 d\phi_2+d\phi_3\right)^2 \\
\label{cp3m}
&&+\frac{1}{4}\cos^2\psi\left(d\theta_1^2+\sin^2\theta_1
d\phi_1^2\right)+\frac{1}{4}\sin^2\psi\left(d\theta_2^2+\sin^2\theta_2
d\phi_2^2\right).\eea

The membrane embedding into $R_t\times S^7/Z_k$, appropriate for our
purposes, is \bea\nn X_{0}=\frac{R}{2}t(\xi^m), \h W_{a}=R
r_a(\xi^m)e^{i\varphi_a(\xi^m)},\h a=(1,2,3,4),\eea where $t$ is the
$AdS$ time, $r_a$ are real functions of $\xi^m$, while $\varphi_a$
are the isometric coordinates on which the background metric does
not depend. The four complex coordinates $W_{a}$ are restricted by
the real embedding condition \bea\nn
\sum_{a=1}^{4}W_{a}\bar{W}_{a}=R^2,\h \mbox{or}\h
\sum_{a=1}^{4}r_a^2=1.\eea The coordinates $r_a$ are connected to
the initial coordinates, on which the background depends, in an
obvious way.

For the embedding described above, the metric induced on the
M2-brane worldvolume is given by \bea\label{im}
G_{mn}=\frac{R^2}{4}\left[-\p_mt\p_nt+
4\sum_{a=1}^{4}\left(\p_mr_a\p_nr_a +
r_a^2\p_m\varphi_a\p_n\varphi_a\right)\right].\eea Correspondingly,
the membrane Lagrangian becomes \bea\nn
\mathcal{L}=\mathcal{L}_{M}+\Lambda\left(\sum_{a=1}^{4}r_a^2-1\right),\eea
where $\Lambda$ is a Lagrange multiplier.

\setcounter{equation}{0}
\section{NR integrable system for M2-branes on $R_t\times S^7/Z_k$}

Let us consider the following particular case of the above membrane
embedding \cite{MNR} \bea\label{nra}
&&X_{0}=\frac{R}{2}\kappa\tau,\h
W_{a}=Rr_a(\xi,\eta)e^{i\left[\omega_{a}\tau+\mu_a(\xi,\eta)\right]},
\\ \nn && \xi=\alpha\sigma_1+\beta\tau,\h \eta=\gamma\sigma_2+\delta\tau,\eea
which implies \bea\label{i} t=\kappa\tau,\h
\varphi_a(\xi^m)=\varphi_a(\tau,\sigma_1,\sigma_2)=
\omega_{a}\tau+\mu_a(\xi,\eta).\eea Here $\kappa$, $\omega_{a}$,
$\alpha$, $\beta$, $\gamma$, $\delta$ are parameters. For this
ansatz, the membrane Lagrangian takes the form ($\p_\xi=\p/\p\xi$,
$\p_\eta=\p/\p\eta$) \bea\nn &&\mathcal{L}=-\frac{R^2}{4\lambda^0}
\left\{\left(2\lambda^0T_2R\alpha\gamma\right)^2
\sum_{a<b=1}^{4}\left[(\p_\xi r_a\p_\eta r_b-\p_\eta r_a\p_\xi r_b)^2\right. \right. \\
\nn &&+ \left. \left. (\p_\xi r_a\p_\eta\mu_b-\p_\eta
r_a\p_\xi\mu_b)^2r_b^2 + (\p_\xi\mu_a\p_\eta
r_b-\p_\eta\mu_a\p_\xi r_b)^2r_a^2\right.\right.
\\ \nn &&+\left.\left.(\p_\xi\mu_a\p_\eta\mu_b-\p_\eta\mu_a\p_\xi\mu_b)^2r_a^2r_b^2 \right]\right. \\
\nn &&+
\left.\sum_{a=1}^{4}\left[\left(2\lambda^0T_2R\alpha\gamma\right)^2
(\p_\xi r_a\p_\eta\mu_a-\p_\eta r_a\p_\xi\mu_a)^2-
\left(\beta\p_\xi\mu_a+\delta\p_\eta\mu_a+\omega_a\right)^2\right]r_a^2\right.
\\
\nn &&-\left.\sum_{a=1}^{4}\left(\beta\p_\xi r_a+\delta\p_\eta
r_a\right)^2+(\kappa/2)^2\right\}+\Lambda\left(\sum_{a=1}^{4}r_a^2-1\right).\eea

We have found a set of sufficient conditions, which reduce the above
Lagrangian to the NR one. First of all, two of the angles
$\varphi_a$ should be set to zero. The corresponding $r_a$
coordinates must depend only on $\eta$ in a specific way. The
remaining variables $r_a$ and $\mu_a$ can depend only on
$\xi$\footnote{Of course, the roles of $\xi$ and $\eta$ can be
interchanged in this context.}. In principle, there are six such
possibilities. How they are realized for the $R_t\times S^7/Z_k$
background, we will discuss in the next section. Here, we will work
out the following example \bea\nn &&r_{1}=r_{1}(\xi),\h
r_{2}=r_{2}(\xi),\h \mu_1=\mu_1(\xi),\h \mu_2=\mu_2(\xi),\\
\label{NRA}
&&r_3=r_3(\eta)=r_0\sin\eta,\h r_4=r_4(\eta)=r_0\cos\eta,\h r_0<1,\\
\nn &&\varphi_3=\varphi_4=0.\eea For this choice, we receive (prime
is used for $d/d\xi$) \bea\nn &&\mathcal{L}=-\frac{R^2}{4\lambda^0}
\left\{\sum_{a=1}^{2}\left[(\tilde{A}^2-\beta^2)r_a'^2\right.+
\left.(\tilde{A}^2-\beta^2)r_a^2\left(\mu'_a-\frac{\beta\omega_a}{\tilde{A}^2-\beta^2}\right)^2
- \frac{\tilde{A}^2}{\tilde{A}^2-\beta^2}\omega_a^2 r_a^2\right]\right. \\
\nn &&+\left. (\kappa/2)^2-r_0^2\delta^2\right\} +
\Lambda\left[\sum_{a=1}^{2}r_a^2-(1-r_0^2)\right],\eea where $
\tilde{A}^2\equiv \left(2\lambda^0T_2R\alpha\gamma r_0\right)^2$.
Now we can integrate once the equations of motion for $\mu_a$
following from the above Lagrangian to get \bea\label{mus}
\mu'_a=\frac{1}{\tilde{A}^2-\beta^2}\left(\frac{C_a}{r_a^2}+\beta\omega_a\right),\eea
where $C_a$ are arbitrary constants. By using (\ref{mus}) in the
equations of motion for $r_a(\xi)$, one finds that they can be
obtained from the effective Lagrangian \bea\nn L_{NR}=
\sum_{a=1}^{2}\left[(\tilde{A}^2-\beta^2)r_a'^2 -
\frac{1}{\tilde{A}^2-\beta^2}\left(\frac{C_a^2}{r_a^2} +
\tilde{A}^2\omega_a^2
r_a^2\right)\right]+\Lambda_M\left[\sum_{a=1}^{2}r_a^2-(1-r_0^2)\right].\eea
This Lagrangian, in full analogy with the string considerations
\cite{AFRT,ART,KRT06}, corresponds to particular case of the
$n$-dimensional NR integrable system. For $C_a=0$ one obtains the
Neumann integrable model, which in the case at hand describes
two-dimensional harmonic oscillator, constrained to a circle of
radius $\sqrt{1-r_0^2}$.

Let us consider the three constraints (\ref{00gf}), (\ref{0igf}) for
the present case. For more close correspondence with the string
case, we want the third one, $G_{02}=0$, to be identically
satisfied. To this end, since $G_{02}\sim r_0^2\gamma\delta,$ we set
$\delta=0$, i.e. $\eta=\gamma\sigma_2$. Then, the first two
constraints give the conserved Hamiltonian $H_{NR}$ and a relation
between the parameters involved: \bea\nn
&&H_{NR}=\sum_{a=1}^{2}\left[(\tilde{A}^2-\beta^2)r_a'^2+
\frac{1}{\tilde{A}^2-\beta^2}\left(\frac{C_a^2}{r_a^2} +
\tilde{A}^2\omega_a^2
r_a^2\right)\right]=\frac{\tilde{A}^2+\beta^2}{\tilde{A}^2-\beta^2}
(\kappa/2)^2,
\\ \label{effcs}&&\sum_{a=1}^{2}\omega_{a}C_a +
\beta(\kappa/2)^2=0.\eea

For closed membranes, $r_a$ and $\mu_a$ must satisfy the following
periodicity conditions \bea
r_a(\xi+2\pi\alpha,\eta+2\pi\gamma)=r_a(\xi,\eta),\h
\mu_a(\xi+2\pi\alpha,\eta+2\pi\gamma)=\mu_a(\xi,\eta)+2\pi
n_a,\label{pbc} \eea where $n_a$ are integer winding numbers. In
particular, $\gamma$ is a non-zero integer.

Since the background metric does not depend on $t$ and $\varphi_a$,
the corresponding conserved quantities are the membrane energy $E$
and four angular momenta $J_a$, defined by \bea\nn E=-\int
d^2\sigma\frac{\p\mathcal{L}}{\p(\p_0 t)},\h J_a=\int
d^2\sigma\frac{\p\mathcal{L}}{\p(\p_0\varphi_a)},\h a=1,2,3,4.\eea
For our ansatz (\ref{NRA}) $J_3=J_4=0$. The energy and the other two
angular momenta are given by \bea\label{cqs} E=\frac{\pi
R^2\kappa}{4\lambda^0\alpha}\int d\xi,\h J_a=\frac{\pi
R^2}{\lambda^0\alpha(\tilde{A}^2-\beta^2)}\int d\xi \left(\beta C_a
+ \tilde{A}^2\omega_a r_a^2\right),\h a=1,2.\eea From here, by using
the constraints (\ref{effcs}), one obtains the energy-charge
relation \bea\nn
\frac{4}{\tilde{A}^2-\beta^2}\left[\tilde{A}^2(1-r_0^2) +
\beta\sum_{a=1}^{2}\frac{C_a}{\omega_a}\right]\frac{E}{\kappa}
=\sum_{a=1}^{2}\frac{J_a}{\omega_a}.\eea As usual, it is linear with
respect to $E$ and $J_a$ before taking the semiclassical limit.

To identically satisfy the embedding condition \bea\nn
\sum_{a=1}^{2}r_a^2-(1-r_0^2)=0,\eea we set \bea\nn
r_1(\xi)=\sqrt{1-r_0^2}\sin\theta(\xi),\h
r_2(\xi)=\sqrt{1-r_0^2}\cos\theta(\xi).\eea Then from the
conservation of the NR Hamiltonian (\ref{effcs}) one finds
\bea\label{mtsol} &&\theta'=\frac{\pm 1}{\tilde{A}^2-\beta^2}
\left[(\tilde{A}^2+\beta^2)\tilde{\kappa}^2 -
\frac{\tilde{C}_1^2}{\sin^2{\theta}} -
\frac{\tilde{C}_2^2}{\cos^2{\theta}} -
\tilde{A}^2\left(\omega_1^2\sin^2{\theta}
+\omega_2^2\cos^2{\theta}\right)\right]^{1/2},\\ \nn
&&\sum_{a=1}^{2}\omega_{a}\tilde{C}_a + \beta\tilde{\kappa}^2=0,\h
\tilde{\kappa}^2=\frac{(\kappa/2)^2}{1-r_0^2},\h
\tilde{C}_a^2=\frac{C_a^2}{(1-r_0^2)^2}.\eea By replacing the
solution for $\theta(\xi)$ received from (\ref{mtsol}) into
(\ref{mus}), one obtains the solutions for $\mu_a$:
\bea\label{mu12s}
\mu_1=\frac{1}{\tilde{A}^2-\beta^2}\left(\tilde{C}_1\int\frac{d\xi}{\sin^2{\theta}}
+ \beta\omega_1\xi\right),\h
\mu_2=\frac{1}{\tilde{A}^2-\beta^2}\left(\tilde{C}_2\int\frac{d\xi}{\cos^2{\theta}}
+ \beta\omega_2\xi\right).\eea

The above analysis shows that the NR integrable models for membranes
on $R_t\times S^7$ and $R_t\times S^7/Z_k$ are the same \cite{AB2}.
Therefore, we can use the results obtained in \cite{AB2} for the
present case. For convenience, the corresponding solutions and
dispersion relations are given in the appendix.

\setcounter{equation}{0}
\section{M2-brane solutions on $R_t\times S^7/Z_k$}

For our membrane embedding in $R_t\times S^7/Z_k$, the angular
variables $\varphi_a$ are related to the corresponding background
coordinates as follows \bea\nn
&&\varphi_1=\frac{\varphi}{k}+\frac{1}{2}\left(\phi_1+\phi_3\right),
\h\varphi_2=\frac{\varphi}{k}-\frac{1}{2}\left(\phi_1-\phi_3\right),
\\ \nn &&\varphi_3=\frac{\varphi}{k}+\frac{1}{2}\left(\phi_2-\phi_3\right),
\h\varphi_4=\frac{\varphi}{k}-\frac{1}{2}\left(\phi_2+\phi_3\right).\eea
As a consequence, for the angular momenta we have \bea\nn
&&J_{\varphi_1}=\frac{J_{\varphi}}{k}+\frac{1}{2}\left(J_{\phi_1}+J_{\phi_3}\right),
\h
J_{\varphi_2}=\frac{J_{\varphi}}{k}-\frac{1}{2}\left(J_{\phi_1}-J_{\phi_3}\right),
\\ \nn && J_{\varphi_3}=\frac{J_{\varphi}}{k}+\frac{1}{2}\left(J_{\phi_2}-J_{\phi_3}\right),
\h
J_{\varphi_4}=\frac{J_{\varphi}}{k}-\frac{1}{2}\left(J_{\phi_2}+J_{\phi_3}\right).\eea
$\varphi_a$ and $J_{\varphi_a}$ satisfy the equalities \bea\nn
\sum_{a=1}^{4}\varphi_a=\frac{4}{k}\varphi,\h
\sum_{a=1}^{4}J_{\varphi_a}=\frac{4}{k}J_{\varphi}.\eea

One of the conditions for the existence of NR description of the
M2-brane dynamics is that two of the angles $\varphi_a$ must be
zero, which means that two of the four angular momenta
$J_{\varphi_a}$ vanish. The six possible cases are
\bea\nn
&&\bullet\ \varphi_1=\phi_3+\frac{\phi_1}{2}=\frac{2}{k}\varphi+\frac{\phi_1}{2},\h
\varphi_2=\phi_3-\frac{\phi_1}{2}=\frac{2}{k}\varphi-\frac{\phi_1}{2},
\h \varphi_3=0,\h \varphi_4=0; \\ \nn
&&\bullet\ \varphi_1=\phi_1=\frac{2}{k}\varphi+\phi_3,\h
\varphi_3=\phi_2=\frac{2}{k}\varphi-\phi_3,\h \varphi_2=0,\h
\varphi_4=0; \\ \label{spc}
&&\bullet\ \varphi_1=\phi_1=\frac{2}{k}\varphi+\phi_3,\h
\varphi_4=-\phi_2=\frac{2}{k}\varphi-\phi_3,\h \varphi_2=0,\h
\varphi_3=0;
\\ \nn &&\bullet\ \varphi_2=-\phi_1=\frac{2}{k}\varphi+\phi_3,\h \varphi_3=\phi_2=\frac{2}{k}\varphi-\phi_3,\h
\varphi_1=0,\h \varphi_4=0;
\\ \nn &&\bullet\ \varphi_2=-\phi_1=\frac{2}{k}\varphi+\phi_3,\h \varphi_4=-\phi_2=\frac{2}{k}\varphi-\phi_3,\h
\varphi_1=0,\h \varphi_3=0; \\ \nn &&\bullet\
\varphi_3=-\phi_3+\frac{\phi_2}{2}=\frac{2}{k}\varphi+\frac{\phi_2}{2},\h
\varphi_4=-\phi_3-\frac{\phi_2}{2}=\frac{2}{k}\varphi-\frac{\phi_2}{2},
\h \varphi_1=0,\h \varphi_2=0 .\eea Here, $\phi_1$ and $\phi_2$ are
the isometry angles on the two two-spheres inside $\mathbb {CP}^3$,
while $\phi_3$ is isometry angle on the $U(1)$ fiber over $S^2\times
S^2$, as can be seen from (\ref{cp3m}).

From (\ref{spc}) it is clear that we have two alternative
descriptions for $\varphi_a$. One is only in terms of the isometry
angles on $\mathbb {CP}^3$, and the other includes the eleventh
coordinate $\varphi$. This is a consequence of our restriction to
M2-brane configurations, which can be described by the NR integrable
system.

The six cases above can be divided into two classes. The first one
contains the first and last possibilities, and the other one - the
remaining ones.
The cases belonging to the first class are related to each other by
the exchange of $\phi_1$ and $\phi_2$. This corresponds to
exchanging the two $S^2$ inside $\mathbb {CP}^3$. Since these
spheres enter symmetrically, the two cases are equivalent.
In terms of ($\varphi$,$\phi_3$), the four cases from the second
class are actually identical. That is why, all of them can be
described simultaneously by choosing one representative from the
class.

Let us first give the M2-brane solutions for cases in the first class.
Since they correspond to our example in the previous section, the
membrane configuration reads \bea\nn
&&W_1=Rr_1(\xi)\exp\left\{i\varphi_1(\tau,\xi)
\right\}=R\sqrt{1-r_0^2}\sin\theta(\xi)\exp\left\{i\left[\frac{2}{k}\varphi(\tau,\xi)
+ \frac{\phi(\tau,\xi)}{2}\right]\right\},\\ \nn
&&W_2=Rr_2(\xi)\exp\left\{i\varphi_2(\tau,\xi)
\right\}=R\sqrt{1-r_0^2}\cos\theta(\xi)\exp\left\{i\left[\frac{2}{k}\varphi(\tau,\xi)
-\frac{\phi(\tau,\xi)}{2}\right]\right\},\\ \nn
&&W_3=Rr_0\sin(\gamma\sigma_2),\h W_4=Rr_0\cos(\gamma\sigma_2),\eea
where $\phi$ is equal to $\phi_1$ or $\phi_2$.

From the NR system viewpoint, the membrane solutions for the second
class configurations differ from the ones just given by the exchange
of $W_2$, $W_3$, and by the replacement $\phi/2\to\phi_3$. In other
words, we have \bea\nn
&&W_1=Rr_1(\xi)\exp\left\{i\varphi_1(\tau,\xi)
\right\}=R\sqrt{1-r_0^2}\sin\theta(\xi)\exp\left\{i\left[\frac{2}{k}\varphi(\tau,\xi)
+ \phi_3(\tau,\xi)\right]\right\},\\ \nn
&&W_2=Rr_0\sin(\gamma\sigma_2),\\ \nn
&&W_3=Rr_3(\xi)\exp\left\{i\varphi_3(\tau,\xi)
\right\}=R\sqrt{1-r_0^2}\cos\theta(\xi)\exp\left\{i\left[\frac{2}{k}\varphi(\tau,\xi)
-\phi_3(\tau,\xi)\right]\right\},\\ \nn
&&W_4=Rr_0\cos(\gamma\sigma_2),\eea

The explicit solutions for $\theta(\xi)$ and $\varphi_{1,2,3}(\tau,\xi)$, of
the M2-brane GM and SS, along with the energy-charge relations
for the infinite and finite sizes are given in the appendix.
Here, we will present them in terms of $\varphi$ and $\phi_{1,2,3}$.

In accordance with (\ref{EJGM0}), we have for the M2-brane GM with
two angular momenta the following dispersion relation
\bea\label{gm0}
\sqrt{1-r_0^2}E-\frac{1}{2}\left(\frac{2}{k}J_{\varphi}+J_{\phi}\right)
=\sqrt{\frac{1}{4}\left(\frac{2}{k}J_{\varphi}-J_{\phi}\right)^2 +
8\lambda k^2\left[r_0(1-r_0^2)\gamma\right]^2\sin^2\frac{p}{2}},\eea
where $J_{\phi}$ can be equal to $J_{\phi_1}/2$, $J_{\phi_2}/2$ or
$J_{\phi_3}$. In writing (\ref{gm0}), we have used that \bea\nn
R=l_p\left(2^5\pi^2kN\right)^{1/6},\h
T_2=\frac{1}{(2\pi)^2l_p^3},\eea and the 't Hooft coupling is
defined by $\lambda=N/k$.

If we introduce the notations \bea\label{mnot}
\mathcal{E}=a\sqrt{1-r_0^2}E,\h
\mathcal{J}_{\varphi}=a\frac{J_{\varphi}}{2},\h
\mathcal{J}_{\phi}=a\frac{J_{\phi}}{2},\h
a=\frac{1}{\sqrt{2\lambda}kr_0(1-r_0^2)\gamma},\eea the above
energy-charge relation takes the form \bea\nn
\mathcal{E}-\mathcal{J}_1(k) =\sqrt{\mathcal{J}^2_2(k) +
4\sin^2\frac{p}{2}},\eea where \bea\label{Jnot}
\mathcal{J}_1(k)=\frac{2}{k}\mathcal{J}_{\varphi}+\mathcal{J}_{\phi},\h
\mathcal{J}_2(k)=\frac{2}{k}\mathcal{J}_{\varphi}-\mathcal{J}_{\phi}.\eea

By using (\ref{mnot}), (\ref{Jnot}) and (\ref{IEJ1}), we can write
down the dispersion relation for the dyonic GM, including the
leading finite-size correction as \bea\nn
&&\mathcal{E}-\mathcal{J}_1(k) =\sqrt{\mathcal{J}^2_2(k) +
4\sin^2\frac{p}{2}} - \frac{16 \sin^4\frac{p}{2}}
{\sqrt{\mathcal{J}^2_2(k)+4\sin^2\frac{p}{2}}}\\
\nn &&\exp\left[-\frac{2\sin^2\frac{p}{2}\left(\mathcal{J}_1(k) +
\sqrt{\mathcal{J}^2_2(k)+4\sin^2\frac{p}{2}}\right)
\sqrt{\mathcal{J}^2_2(k) +4\sin^2\frac{p}{2}}}{\mathcal{J}^2_2(k)
+4\sin^4\frac{p}{2}}\right].\eea

The reason to introduce $\mathcal{E}$, $\mathcal{J}_{\varphi}$ and
$\mathcal{J}_{\phi}$ namely in this way is the following. For GM on
the $R_t\times S^3$ subspace of $AdS_5\times S^5$, in terms of
\bea\nn \mathcal{E}=\frac{2\pi}{\sqrt{\lambda}}E,\h
\mathcal{J}_{1}=\frac{2\pi}{\sqrt{\lambda}}J_1,\h
\mathcal{J}_{2}=\frac{2\pi}{\sqrt{\lambda}}J_2,\eea we have \bea\nn
\mathcal{E}-\mathcal{J}_{1} =\sqrt{\mathcal{J}_{2}^2 +
4\sin^2\frac{p}{2}}.\eea The same result can be obtained for the GM
on the $R_t\times \mathbb {CP}^3$ subspace of $AdS_4\times \mathbb
{CP}^3$, if we use the identification \cite{ABR1} \bea\nn
\mathcal{E}=\frac{E}{\sqrt{2\lambda}},\h
\mathcal{J}_{1}=\frac{J_1}{\sqrt{2\lambda}},\h
\mathcal{J}_{2}=\frac{J_2}{\sqrt{2\lambda}}.\eea In the all three
cases, the second term under the square root is the same. In this
description it is universal - for different backgrounds and for
different extended objects.

Analogously, for the SS case one can find (see (\ref{ssS3c}))
\bea\nn &&\mathcal{E}-\Delta\varphi_1=
p+8\sin^2\frac{p}{2}\tan\frac{p}{2}
\exp\left(-\frac{(\Delta\varphi_1+ p)\tan\frac{p}{2}}
{\mathcal{J}^2_2(k) \csc^2p+\tan^2\frac{p}{2}}\right) \\ \nn
&&\mathcal{J}_1(k)= \sqrt{\mathcal{J}^2_2(k)
+4\sin^2\frac{p}{2}}.\eea

Let us point out that for $k=1$ the above dispersion relations
coincide with the ones obtained earlier in \cite{AB2}. We can also
reproduce the energy-charge relations for dyonic GM and SS strings
on $R_t\times \mathbb {CP}^3$ by taking an appropriate limit. To
show this, let us consider the second case in (\ref{spc}), for which
\bea\nn J_{\phi_1}=\frac{2}{k}J_{\varphi}+J_{\phi_3},\h
J_{\phi_2}=\frac{2}{k}J_{\varphi}-J_{\phi_3}.\eea In accordance with
our membrane embedding, the following identification should be made
\bea\nn J_1^{str}=\frac{J_{\phi_1}}{2},\h
J_2^{str}=\frac{J_{\phi_2}}{2}.\eea Then in the limit $k\to \infty$,
$r_0\to 0$, such that $kr_0\gamma=1$, we obtain from (\ref{gm0})
\bea\nn E-J_1^{str}=\sqrt{(J_2^{str})^2 + 8\lambda
\sin^2\frac{p}{2}}.\eea This is exactly what we have derived in
\cite{ABR1} for dyonic GM strings on $R_t\times \mathbb {CP}^3$.
Obviously, this also applies for the leading finite-size correction.
In the same way, the SS string dispersion relation for $R_t\times
\mathbb {CP}^3$ background can be reproduced.

\setcounter{equation}{0}
\section{Concluding Remarks}
We presented here an M2-brane perspective on ABJM super
Chern-Simons-matter theory, which for large $N$ and finite level $k$
is dual to M theory on $AdS_4\times S^7/Z_k$. In particular, we
found membrane configurations, which can be described by the same NR
integrable model for any positive integer $k$. Based on this, we gave
the explicit solutions and the dispersion relations, including
finite-size corrections, for states with two angular momenta,
exhibiting GM and SS properties.

It would be interesting to see if the above results could be
generalized to M2-branes with three and four angular momenta. Also,
one can try to include the conserved spin $S$, arising from the
nontrivial dynamics on the $AdS_4$ part of the background. In this
case, one must take into account the membrane interaction with the
three-form gauge field.

\section*{Acknowledgements}
This work was supported in part by KRF-2007-313-C00150 (CA), by
NSFB VU-F-201/06 (PB), and by the Brain Pool program from the
Korean Federation of Science and Technology (2007-1822-1-1).

\def\theequation{A.\arabic{equation}}
\setcounter{equation}{0}
\begin{appendix}

\section{M2-brane GM and SS}

For the GM-like case by using that  $\tilde{C}_2=0$,
$\tilde{\kappa}^2=\omega_1^2$ in (\ref{mtsol}), (\ref{mu12s}), one
finds \bea\nn
&&\cos\theta(\xi)=\frac{\cos\tilde{\theta}_0}{\cosh\left(D_0\xi\right)},\h
\sin^2\tilde{\theta}_0=\frac{\beta^2\omega_1^2}{\tilde{A}^2(\omega_1^2-\omega_2^2)},\h
D_0=\frac{\tilde{A}\sqrt{\omega_1^2-\omega_2^2}}
{\tilde{A}^2-\beta^2}\cos\tilde{\theta}_0,
\\ \nn &&\varphi_1(\tau,\xi)=\omega_1\tau +\arctan\left[\cot\tilde{\theta}_0\tanh(D_0\xi)\right],
\h\varphi_2(\tau,\xi)=\omega_2\left(\tau +
\frac{\beta}{\tilde{A}^2-\beta^2}\xi\right).\eea

For the SS-like solutions when $\tilde{C}_2=0$,
$\tilde{\kappa}^2=\omega_1^2\tilde{A}^2/\beta^2$, by solving the
equations (\ref{mtsol}), (\ref{mu12s}), one arrives at \bea\nn
&&\cos\theta(\xi)=\frac{\cos\tilde{\theta}_1}{\cosh\left(D_1\xi\right)},\h
\sin^2\tilde{\theta}_1=\frac{\tilde{A}^2\omega_1^2}{\beta^2(\omega_1^2-\omega_2^2)},\h
D_1=\frac{\tilde{A}\sqrt{\omega_1^2-\omega_2^2}}
{\tilde{A}^2-\beta^2}\cos\tilde{\theta}_1,
\\ \nn &&\varphi_1(\tau,\xi)=\omega_1\left(\tau
-\frac{\xi}{\beta}\right)
-\arctan\left[\cot\tilde{\theta}_1\tanh(D_1\xi)\right],
\h\varphi_2(\tau,\xi)=\omega_2\left(\tau +
\frac{\beta}{\tilde{A}^2-\beta^2}\xi\right) .\eea

The energy-charge relations computed on the above membrane solutions
were found in \cite{BR07}, and in our notations read
\bea\label{EJGM0} \sqrt{1-r_0^2}E-\frac{J_1}{2}
=\sqrt{\left(\frac{J_2}{2}\right)^2
+\frac{\tilde{\lambda}}{\pi^2}\sin^2\frac{p}{2}},\h
\frac{p}{2}=\frac{\pi}{2}-\tilde{\theta}_0, \eea for the GM-like
case, and \bea\label{EJSS0}
\sqrt{1-r_0^2}E-\frac{\sqrt{\tilde{\lambda}}}{2\pi}\Delta\varphi_1
=\frac{\sqrt{\tilde{\lambda}}}{\pi}\frac{p}{2}, \h \frac{J_1}{2}
=\sqrt{\left(\frac{J_2}{2}\right)^2
+\frac{\tilde{\lambda}}{\pi^2}\sin^2\frac{p}{2}},\h
\frac{p}{2}=\frac{\pi}{2}-\tilde{\theta}_1, \eea for the SS-like
solution, where \bea\label{tl} \tilde{\lambda}=\left[2\pi^2T_2R^3r_0
(1-r_0^2)\gamma\right]^2.\eea

\subsection{Finite-Size Effects}

For $\tilde{C}_2=0$, Eq.(\ref{mtsol}) can be written as
\bea\label{tS3eq}
(\cos\theta)'=\mp\frac{\tilde{A}\sqrt{\omega_1^2-\omega_2^2}}{\tilde{A}^2-\beta^2}
\sqrt{(z_+^2-\cos^2\theta)(\cos^2\theta-z_-^2)},\eea where \bea\nn
&&z^2_\pm=\frac{1}{2(1-\frac{\omega_2^2}{\omega_1^2})}
\left\{q_1+q_2-\frac{\omega_2^2}{\omega_1^2}
\pm\sqrt{(q_1-q_2)^2-\left[2\left(q_1+q_2-2q_1
q_2\right)-\frac{\omega_2^2}{\omega_1^2}\right]
\frac{\omega_2^2}{\omega_1^2}}\right\}, \\ \nn
&&q_1=1-\tilde{\kappa}^2/\omega_1^2,\h
q_2=1-\beta^2\tilde{\kappa}^2/\tilde{A}^2\omega_1^2 .\eea The
solution of (\ref{tS3eq}) is \bea\label{tS3sol} \cos\theta=z_+
dn\left(C\xi|m\right),\h
C=\mp\frac{\tilde{A}\sqrt{\omega_1^2-\omega_2^2}}{\tilde{A}^2-\beta^2}
z_+,\h m\equiv 1-z^2_-/z^2_+ .\eea The solutions of
Eqs.(\ref{mu12s}) now read \bea\nn
&&\mu_1=\frac{2\beta/\tilde{A}}{z_+\sqrt{1-\omega_2^2/\omega_1^2}}
\left[C\xi -
\frac{\tilde{\kappa}^2/\omega_1^2}{1-z^2_+}\Pi\left(am(C\xi),-\frac{z^2_+
-z^2_-}{1-z^2_+}\bigg \vert m\right)\right],\\ \nn
&&\mu_2=\frac{2\beta\omega_2/\tilde{A}\omega_1}{z_+\sqrt{1-\omega_2^2/\omega_1^2}}
C\xi,\eea where $\Pi(k,n|m)$ is the elliptic integral of third kind.

Our next task is to find out what kind of energy-charge relations
can appear for the M2-brane solution in the limit when the energy
$E\to \infty$. It turns out that the semiclassical behavior depends
crucially on the sign of the difference $\tilde{A}^2-\beta^2$.

\subsubsection{The M2-brane GM}

We begin with the M2-brane GM, i.e. $\tilde{A}^2>\beta^2$. In this
case, one obtains from (\ref{cqs}) the following expressions for the
conserved energy $E$ and the angular momenta $J_1$, $J_2$ \bea\nn
&&\mathcal{E} =\frac{2\tilde{\kappa}(1-\beta^2/\tilde{A}^2)}
{\omega_1 z_+\sqrt{1-\omega_2^2/\omega_1^2}}\mathbf{K}
\left(1-z^2_-/z^2_+\right), \\ \label{cqsGM} &&\mathcal{J}_1=
\frac{2 z_+}{\sqrt{1-\omega_2^2/\omega_1^2}} \left[
\frac{1-\beta^2\tilde{\kappa}^2/\tilde{A}^2\omega_1^2}{z^2_+}\mathbf{K}
\left(1-z^2_-/z^2_+\right)-\mathbf{E}
\left(1-z^2_-/z^2_+\right)\right], \\ \nn &&\mathcal{J}_2= \frac{2
z_+ \omega_2/\omega_1 }{\sqrt{1-\omega_2^2/\omega_1^2}}\mathbf{E}
\left(1-z^2_-/z^2_+\right).\eea Here, we have used the notations
\bea\label{not} \mathcal{E}=\frac{2\pi}{\sqrt{\tilde{\lambda}}}
\sqrt{1-r_0^2} E ,\h
\mathcal{J}_1=\frac{2\pi}{\sqrt{\tilde{\lambda}}}\frac{J_1}{2}, \h
\mathcal{J}_2=\frac{2\pi}{\sqrt{\tilde{\lambda}}}\frac{J_2}{2},\eea
where $\tilde{\lambda}$ is defined in (\ref{tl}). The computation of
$\Delta\varphi_1$ gives \bea\label{pws} p\equiv\Delta\varphi_1 &=&
2\int_{\theta_{min}}^{\theta_{max}}\frac{d \theta}{\theta'}\mu'_1=
\\ \nn &-&\frac{2\beta/\tilde{A}}{z_+\sqrt{1-\omega_2^2/\omega_1^2}}
\left[\frac{\tilde{\kappa}^2/\omega_1^2}{1-z^2_+}\Pi\left(-\frac{z^2_+
- z^2_-}{1-z^2_+}\bigg\vert 1-z^2_-/z^2_+\right) -\mathbf{K}
\left(1-z^2_-/z^2_+\right)\right].\eea In the above expressions,
$\mathbf{K}(m)$, $\mathbf{E}(m)$ and $\Pi(n|m)$ are the complete
elliptic integrals.

Expanding the elliptic integrals about $z_-^2=0$, one arrives at
\bea\label{IEJ1} &&\mathcal{E}-\mathcal{J}_1 =
\sqrt{\mathcal{J}_2^2+4\sin^2(p/2)} - \frac{16 \sin^4(p/2)}
{\sqrt{\mathcal{J}_2^2+4\sin^2(p/2)}}\\ \nn
&&\exp\left[-\frac{2\left(\mathcal{J}_1 +
\sqrt{\mathcal{J}_2^2+4\sin^2(p/2)}\right)
\sqrt{\mathcal{J}_2^2+4\sin^2(p/2)}\sin^2(p/2)}{\mathcal{J}_2^2+4\sin^4(p/2)}
\right].\eea It is easy to check that the energy-charge relation
(\ref{IEJ1}) coincides with the one found in \cite{HS08}, describing
the finite-size effects for dyonic GM. The difference is that in the
string case the relations between $\mathcal{E}$, $\mathcal{J}_1$,
$\mathcal{J}_2$ and $E$, $J_1$, $J_2$ are given by \bea\nn
\mathcal{E}=\frac{2\pi}{\sqrt{\lambda}}E ,\h
\mathcal{J}_1=\frac{2\pi}{\sqrt{\lambda}}J_1, \h
\mathcal{J}_2=\frac{2\pi}{\sqrt{\lambda}}J_2,\eea while for the
M2-brane they are written in (\ref{not}).

\subsubsection{The M2-brane SS}

Let us turn our attention to the M2-brane SS, when
$\tilde{A}^2<\beta^2$. The computation of the conserved quantities
(\ref{cqs}) and $\Delta\varphi_1$ now gives \bea\nn &&\mathcal{E}
=\frac{2\tilde{\kappa}(\beta^2/\tilde{A}^2-1)}
{\omega_1\sqrt{1-\omega_2^2/\omega_1^2}z_+}\mathbf{K}
\left(1-z^2_-/z^2_+\right), \\ \nn &&\mathcal{J}_1= \frac{2
z_+}{\sqrt{1-\omega_2^2/\omega_1^2}} \left[\mathbf{E}
\left(1-z^2_-/z^2_+\right)
-\frac{1-\beta^2\tilde{\kappa}^2/\tilde{A}^2\omega_1^2}{z^2_+}\mathbf{K}
\left(1-z^2_-/z^2_+\right)\right], \\ \nn &&\mathcal{J}_2= -\frac{2
z_+ \omega_2/\omega_1 }{\sqrt{1-\omega_2^2/\omega_1^2}}\mathbf{E}
\left(1-z^2_-/z^2_+\right), \\ \nn &&\Delta\varphi_1=
-\frac{2\beta/\tilde{A}}{\sqrt{1-\omega_2^2/\omega_1^2}z_+}
\left[\frac{\tilde{\kappa}^2/\omega_1^2}{1-z^2_+}\Pi\left(-\frac{z^2_+
- z^2_-}{1-z^2_+}|1-z^2_-/z^2_+\right) -\mathbf{K}
\left(1-z^2_-/z^2_+\right)\right].\eea

$\mathcal{E}-\Delta\varphi_1$ can be derived as
\bea\label{DiffJ12}\mathcal{E}-\Delta\varphi_1&=& \arcsin
N(\mathcal{J}_1,\mathcal{J}_2) +
2\left(\mathcal{J}_1^2-\mathcal{J}_2^2\right)
\sqrt{\frac{4}{\left[4-\left(\mathcal{J}_1^2-\mathcal{J}_2^2\right)\right]}-1}\\
\nn
&\times&\exp\left[-\frac{2\left(\mathcal{J}_1^2-\mathcal{J}_2^2\right)
N(\mathcal{J}_1,\mathcal{J}_2)}
{\left(\mathcal{J}_1^2-\mathcal{J}_2^2\right)^2 +
4\mathcal{J}_2^2}\left[\Delta\varphi +\arcsin
N(\mathcal{J}_1,\mathcal{J}_2)\right]\right],\\
\nn N(\mathcal{J}_1,\mathcal{J}_2)&\equiv&
\frac{1}{2}\left[4-\left(\mathcal{J}_1^2-\mathcal{J}_2^2\right)\right]
\sqrt{\frac{4}{\left[4-\left(\mathcal{J}_1^2-\mathcal{J}_2^2\right)\right]}-1}.\eea
Finally, by using the SS relation between the angular momenta
\bea\nn \mathcal{J}_1=\sqrt{\mathcal{J}_2^2+4\sin^2(p/2)},\eea we
obtain \bea\label{ssS3c} \mathcal{E}-\Delta\varphi_1=
p+8\sin^2\frac{p}{2}\tan\frac{p}{2}
\exp\left(-\frac{\tan\frac{p}{2}(\Delta\varphi_1 + p)}
{\tan^2\frac{p}{2} + \mathcal{J}_2^2 \csc^2p}\right).\eea This
result coincides with the string result found in \cite{AB1}. As in
the GM case, the difference is in the identification (\ref{not}).

\end{appendix}


\begin{thebibliography}{99}

\bibitem{AdS/CFT} J. M. Maldacena, ``The large N limit of superconformal
field theories and supergravity,'' Adv. Theor. Math. Phys. {\bf 2},
231 (1998) [{arXiv:hep-th/9711200}].

\bibitem{GKP98}
S. S. Gubser, I. R. Klebanov and A. M. Polyakov, ``Gauge theory
correlators from non-critical string theory,''
Phys. Lett. {\bf B428}, 105 (1998) [{arXiv:hep-th/9802109}].

\bibitem{EW98}
E. Witten, ``Anti-de Sitter space and holography,'' Adv. Theor.
Math. Phys. {\bf 2}, 253 (1998) [{arXiv:hep-th/9802150}].

\bibitem{CFT3}
J.~H.~Schwarz, ``Superconformal Chern-Simons theories,''
JHEP {\bf 0411}, 078 (2004) [arXiv:hep-th/0411077];
J.~Bagger and N.~Lambert, ``Modeling multiple M2's,''
Phys.\ Rev.\  D {\bf 75}, 045020 (2007) [arXiv:hep-th/0611108];
D.~Gaiotto and X.~Yin, ``Notes on superconformal Chern-Simons-matter theories,''
JHEP {\bf 0708}, 056 (2007)   [arXiv:hep-th/0704.3740];
A.~Gustavsson, ``Algebraic structures on parallel M2-branes,''
[arXiv:hep-th/0709.1260]; J.~Bagger and N.~Lambert,
``Gauge Symmetry and Supersymmetry of Multiple M2-Branes,''
Phys.\ Rev.\  D {\bf 77}, 065008 (2008) [arXiv:hep-th/0711.0955];
J.~Bagger and N.~Lambert, ``Comments On Multiple M2-branes,''
JHEP {\bf 0802}, 105 (2008) [arXiv:hep-th/0712.3738];
M.~Van Raamsdonk,
``Comments on the Bagger-Lambert theory and multiple M2-branes,''
JHEP {\bf 0805}, 105 (2008)   [arXiv:0803.3803 [hep-th]];
J.~Distler, S.~Mukhi, C.~Papageorgakis and M.~Van Raamsdonk,
``M2-branes on M-folds,'' JHEP {\bf 0805}, 038 (2008) [arXiv:hep-th/0804.1256];
P.~M.~Ho, Y.~Imamura and Y.~Matsuo, ``M2 to D2 revisited,''
 [arXiv:hep-th/0805.1202];
C.~Krishnan and C.~Maccaferri, ``Membranes on Calibrations,''
[arXiv:hep-th/0805.3125];
K. Hosomichi, K-M. Lee, S. Lee, S. Lee and J. Park,
``N=4 Superconformal Chern-Simons Theories with Hyper and Twisted Hyper Multiplets ,''
[arXiv:hep-th/0805.3662];
J.~Gomis, D.~Rodriguez-Gomez, M.~Van Raamsdonk and H.~Verlinde,
``The Superconformal Gauge Theory on M2-Branes,''
[arXiv:hep-th/0806.0738].


\bibitem{ABJM0806} O. Aharony, O. Bergman, D. L. Jafferis and J. Maldacena,
``${\cal N}=6$ superconformal Chern - Simons - matter theories,
M2-branes and their gravity duals,'' [arXiv:hep-th/0806.1218v3].

\bibitem{06.1519} M. Benna, I. Klebanov, T. Klose and M.
Smedback, ``Superconformal Chern-Simons Theories and $AdS_4/CFT_3$
Correspondence,'' JHEP {\bf 0809} 072 (2008),
[arXiv:hep-th/0806.1519v3].

\bibitem{06.3251} J. Bhattacharya and S. Minwalla,
``Superconformal Indices for ${\cal N}=6$ Chern Simons Theories,''
[arXiv:hep-th/0806.3251v1].

\bibitem{06.3391} T. Nishioka and T. Takayanagi
``On Type IIA Penrose Limit and ${\cal N}=6$ Chern-Simons
Theories,'' JHEP {\bf 0808} 001 (2008), [arXiv:hep-th/0806.3391v4].

\bibitem{06.3498} Y. Honma, S. Iso, Y. Sumitomo and S. Zhang
``Scaling limit of ${\cal N}=6$ superconformal Chern-Simons theories
and Lorentzian Bagger-Lambert theories,'' [arXiv:hep-th/0806.3498v3].

\bibitem{06.3727} Y. Imamura and K. Kimura
``On the moduli space of elliptic Maxwell-Chern-Simons theories,''
[arXiv:hep-th/0806.3727v2].

\bibitem{06.3951} J. A. Minahan and K. Zarembo,
``The Bethe ansatz for superconformal Chern-Simons,'' JHEP {\bf
0809} 040 (2008), [arXiv:hep-th/0806.3951v5].

\bibitem{06.4589} D. Gaiotto, S. Giombi and X. Yin,
``Spin Chains in ${\cal N}=6$ Superconformal Chern-Simons-Matter
Theory,'' [arXiv:hep-th/0806.4589v1].

\bibitem{06.4940} G. Arutyunov and S. Frolov,
``Superstrings on $AdS_4 \times CP^3$ as a Coset Sigma-model,''
[arXiv:hep-th/0806.4940v2].

\bibitem{06.4948} B. Stefanski jr,
``Green-Schwarz action for Type IIA strings on $AdS_4\times CP^3$,''
[arXiv:hep-th/0806.4948v2].

\bibitem{06.4959} G. Grignani, T. Harmark and M. Orselli,
``The $SU(2)\times SU(2)$ sector in the string dual of ${\cal N}=6$
superconformal Chern-Simons theory,'' [arXiv:hep-th/0806.4959v4].

\bibitem{06.4977} K. Hosomichi, K-M. Lee, S. Lee, S. Lee and J. Park,
``${\cal N}=5,6$ Superconformal Chern-Simons Theories and M2-branes
on Orbifolds,'' JHEP {\bf 0809} 002 (2008),
[arXiv:hep-th/0806.4977v2].

\bibitem{07.0163} J. Bagger and N. Lambert,
``Three-Algebras and ${\cal N}=6$ Chern-Simons Gauge Theories,''
[arXiv:hep-th/0807.0163v2].

\bibitem{07.0197} S. Terashima,
``On M5-branes in ${\cal N}=6$ Membrane Action,'' JHEP {\bf 0808}
080 (2008), [arXiv:hep-th/0807.0197v3].

\bibitem{07.0205} G. Grignani, T. Harmark, M. Orselli and G. W.
Semenoff, ``Finite size Giant Magnons in the string dual of ${\cal
N}=6$ superconformal Chern-Simons theory,''
[arXiv:hep-th/0807.0205v2].

\bibitem{07.0437} N. Gromov and P. Vieira,
``The AdS4/CFT3 algebraic curve,''
[arXiv:hep-th/0807.0437v2].

\bibitem{AB2} C. Ahn and P. Bozhilov,
`` Finite-size effects of Membranes on $AdS_4\times S_7$,'' JHEP
{\bf 0808} 054 (2008),
[arXiv:hep-th/0807.0566v2].

\bibitem{07.0777} N. Gromov and P. Vieira,
``The all loop AdS4/CFT3 Bethe ansatz,''
[arXiv:hep-th/0807.0777v2].

\bibitem{07.0802} B. Chen and J-B. Wu,
``Semi-classical strings in $AdS_4\times CP^3$,'' JHEP {\bf 0809}
096 (2008),
[arXiv:hep-th/0807.0802v3].

\bibitem{07.1527} D. Astolfi, V. G. M. Puletti, G. Grignani, T. Harmark and M.
Orselli, ``Finite-size corrections in the $SU(2)\times SU(2)$ sector
of type IIA string theory on $AdS_4\times CP^3$,''
[arXiv:hep-th/0807.1527v3].

\bibitem{07.1924} C. Ahn and R. I. Nepomechie,
``${\cal N}=6$ super Chern-Simons theory $S$-matrix and all-loop
Bethe ansatz equations,'' JHEP {\bf 0809} 010 (2008), [arXiv:hep-th/0807.1924v2].

\bibitem{07.2063} D. Bak and S-J. Rey,
``Integrable Spin Chain in Superconformal Chern-Simons Theory,''
[arXiv:hep-th/0807.2063v3].

\bibitem{07.2559}  B-H. Lee, K. L. Panigrahi and C. Park,
``Spiky Strings on $AdS_4 \times CP^3$,''
[arXiv:hep-th/0807.2559v2].

\bibitem{07.2861} I. Shenderovich,
``Giant magnons in $AdS_4/CFT_3$: dispersion, quantization and
finite--size corrections,''
[arXiv:hep-th/0807.2861v3].

\bibitem{ABR1} C. Ahn, P. Bozhilov and R.C. Rashkov,
``Neumann-Rosochatius integrable system for strings on $AdS_4\times
\mathbb {CP}^3$,''  JHEP {\bf 0809} 017 (2008),
[arXiv:hep-th/0807.3134v2].

\bibitem{07.3965} T. McLoughlin and R. Roiban,
``Spinning strings at one-loop in $AdS_4 \times P^3$,''
[arXiv:hep-th/0807.3965v1].

\bibitem{07.4400} L. F. Alday, G. Arutyunov and D. Bykov,
``Semiclassical Quantization of Spinning Strings in $AdS_4 \times
CP^3$,'' [arXiv:hep-th/0807.4400v1].

\bibitem{07.4561} C. Krishnan,
``AdS4/CFT3 at One Loop,'' JHEP {\bf 0809} 092 (2008),
[arXiv:hep-th/0807.4561v3].

\bibitem{07.4897} N. Gromov and V. Mikhaylov,
``Comment on the Scaling Function in $AdS_4 \times CP^3$,''
[arXiv:hep-th/0807.4897v1].

\bibitem{07.4924} O. Aharony, O. Bergman and D. L. Jafferis,
``Fractional M2-branes,''
[arXiv:hep-th/0807.4924v2].

\bibitem{BT}
D.~Berenstein and D.~Trancanelli,
``Three-dimensional ${\cal N}=6$  SCFT's and their membrane dynamics,''
[arXiv:0808.2503].

\bibitem{08.3057} R.C. Rashkov,
``A note on the reduction of the $AdS_4 \times CP^3$ string sigma
model,'' [arXiv:hep-th/0808.3057v2].

\bibitem{09.1771} K. Hosomichi, K-M. Lee, S. Lee, S. Lee, J. Park and P.
Yi, ``A Nonperturbative Test of M2-Brane Theory,''
[arXiv:hep-th/0809.1771v2].

\bibitem{09.4038} T. McLoughlin, R. Roiban and Arkady A. Tseytlin,
``Quantum spinning strings in $AdS_4 \times CP^3$: testing the Bethe
Ansatz proposal,'' [arXiv:hep-th/0809.4038v1].

\bibitem{09.5106} S. Ryang,
``Giant Magnon and Spike Solutions with Two Spins in $AdS_4 \times
CP^3$,'' [arXiv:hep-th/0809.5106v1].

\bibitem{10.0704}
D.~Bombardelli and D.~Fioravanti, ``Finite-Size Corrections of the
$\mathbb{CP}^3$ Giant Magnons: the L\"{u}scher terms,''
[arXiv:0810.0704].

\bibitem{10.1246}
T.~Lukowski and O.~O.~Sax, ``Finite size giant magnons in the $SU(2)
\times SU(2)$ sector of $AdS_4 \times CP^3$,'' [arXiv:0810.1246].

\bibitem{HM06} D.M. Hofman and J. Maldacena,
``Giant magnons,'' J. Phys. {\bf A 39} 13095 (2006),
[arXiv:hep-th/0604135v2].

\bibitem{ND} N. Dorey,
``Magnon bound states and the AdS/CFT correspondence,'' J. Phys.
{\bf A39}, 13119 (2006) [arXiv:hep-th/0604175];
H.-Y. Chen, N. Dorey and K. Okamura, ``Dyonic giant magnons,''
JHEP {\bf 0609} 024 (2006), [arXiv:hep-th/0605155].

\bibitem{IK07} R. Ishizeki and M. Kruczenski,
``Single spike solutions for strings on $S^2$ and $S^3$,'' Phys.
Rev. {\bf D 76} 126006 (2007) [arXiv:hep-th/0705.2429v2].

\bibitem{NPB656} P. Bozhilov,
``Probe Branes Dynamics: Exact Solutions in General Backgrounds,''
Nucl. Phys. {\bf B656} 199 (2003), [arXiv:hep-th/0211181v3].

\bibitem{MNR} P. Bozhilov,
`` Neumann and Neumann-Rosochatius integrable systems from membranes
on $AdS_4\times S^7$,'' JHEP {\bf 0708} 073 (2007),
[arXiv:hep-th/0704.3082v4].

\bibitem{AFRT}  G. Arutyunov, S. Frolov, J. Russo and A.A. Tseytlin,
``Spinning strings in $AdS_5 \times S^5$ and integrable systems,''
Nucl. Phys. {\bf B671} 3 (2003), [arXiv:hep-th/0307191].

\bibitem{ART} G. Arutyunov, J. Russo and A.A. Tseytlin, ``Spinning strings in
$AdS_5\times S^5$: New integrable system relations,'' Phys. Rev.
{\bf D69} 086009 (2004), [arXiv:hep-th/0311004].

\bibitem{KRT06} M. Kruczenski, J. Russo and A.A. Tseytlin,
``Spiky strings and giant magnons on S5,'' JHEP {\bf 0610} 002
(2006), [arXiv:hep-th/0607044v3].

\bibitem{BR07} P. Bozhilov and R.C. Rashkov,
``On the multi-spin magnon and spike solutions from membranes,''
Nucl. Phys. {\bf B 794} 429 (2008),
[arXiv:hep-th/0708.0325v3]

\bibitem{HS08} Y. Hatsuda and R. Suzuki, ``Finite-Size Effects for Dyonic Giant
Magnons,'' Nucl. Phys. {\bf B 800} 349 (2008),
[arXiv:hep-th/0801.0747v4]/

\bibitem{AB1} C. Ahn and P. Bozhilov,
``Finite-size Effects for Single Spike,'' JHEP {\bf 0807} 105
(2008), [arXiv:hep-th/0806.1085v3].




\end{thebibliography}
\end{document}